\documentclass[dvips]{actaprep}
\usepackage{supertabular,lscape,epsfig}
\usepackage{amssymb}
\usepackage{amsmath}
\DeclareSymbolFont{ppa}{OT1}{ppl}{m}{it}
\DeclareMathSymbol{\vv}{\mathalpha}{ppa}{'166}

\newfont{\hb}{rphvb at 10pt}
\newfont{\hbo}{rphvbo at 10pt}
\newfont{\bitt}{rptmbi at 12pt}
\newfont{\bits}{rptmbi at 11pt}

\SetPages{43}{57}

\SetVol{55}{2005}

\begin{document}

\newcommand{\TabCapp}[2]{\begin{center}\parbox[t]{#1}{\centerline{
  \small {\spaceskip 2pt plus 1pt minus 1pt T a b l e}
  \refstepcounter{table}\thetable}
  \vskip2mm
  \centerline{\footnotesize #2}}
  \vskip3mm
\end{center}}

\newcommand{\TTabCap}[3]{\begin{center}\parbox[t]{#1}{\centerline{
  \small {\spaceskip 2pt plus 1pt minus 1pt T a b l e}
  \refstepcounter{table}\thetable}
  \vskip2mm
  \centerline{\footnotesize #2}
  \centerline{\footnotesize #3}}
  \vskip1mm
\end{center}}

\newcommand{\MakeTableSepp}[4]{\begin{table}[p]\TabCapp{#2}{#3}
  \begin{center} \TableFont \begin{tabular}{#1} #4 
  \end{tabular}\end{center}\end{table}}

\newcommand{\MakeTableee}[4]{\begin{table}[htb]\TabCapp{#2}{#3}
  \begin{center} \TableFont \begin{tabular}{#1} #4
  \end{tabular}\end{center}\end{table}}

\newcommand{\MakeTablee}[5]{\begin{table}[htb]\TTabCap{#2}{#3}{#4}
  \begin{center} \TableFont \begin{tabular}{#1} #5 
  \end{tabular}\end{center}\end{table}}

\newfont{\bb}{ptmbi8t at 12pt}
\newfont{\bbb}{cmbxti10}
\newfont{\bbbb}{cmbxti10 at 9pt}
\newcommand{\uprule}{\rule{0pt}{2.5ex}}
\newcommand{\douprule}{\rule[-2ex]{0pt}{4.5ex}}
\newcommand{\dorule}{\rule[-2ex]{0pt}{2ex}}
\begin{Titlepage}
\Title{The Optical Gravitational Lensing Experiment. Internet Access to
the OGLE Photometry Data Set: OGLE-II {\bitt BVI} maps and
{\bitt I}-band data\footnote{Based on observations obtained with the 1.3~m
Warsaw telescope at the Las Campanas Observatory of the Carnegie
Institution of Washington.}}
\vspace*{-3pt}
\Author{M.\,K.~~S~z~y~m~a~{\'n}~s~k~i}{Warsaw University Observatory,
Al.~Ujazdowskie~4, 00-478~Warszawa, Poland\\ 
e-mail: msz@astrouw.edu.pl}
\vspace*{-3pt}
\Received{March 15, 2005}
\end{Titlepage}
\vspace*{-12pt}
\Abstract{We present on-line, interactive interface to the whole {\it
I}-band photometry data set obtained in the second phase of the OGLE
project (OGLE-II). The raw photometric database is accessed through an
additional database using {\sc MySQL} engine, allowing to select objects
fulfilling any set of criteria including {\it RA/Dec} coordinates, mean
brightness, error etc. The results of the queries can be browsed on-line,
the light curves can be plotted interactively, the photometric data can be
downloaded for the total of over $10^{10}$ measurements of more than 40
million objects in the Galactic bulge and the Magellanic Clouds collected
during OGLE-II. The {\sc MySQL} database of parameters also includes the
complete data set of the previously published photometric {\it BVI} maps of
OGLE-II targets, allowing to interactively select objects from these
maps.}{Catalogs -- Surveys -- Galaxy: bulge -- Magellanic Clouds}

\Section{Introduction}
The data collected during the second phase of the Optical Gravitational
Lensing Experiment (OGLE-II, Udalski, Kubiak and Szyma{\'n}ski 1997)
constitute a unique, extremely rich source of accurate, long baseline,
standard ({\it BVI}) photometry, suitable for many astronomical projects,
not only related to microlensing. The longstanding policy of the OGLE
project has been to make the data available to the wide astronomical
community. Following the publication of the {\it BVI} photometry maps of
the OGLE-II fields (Udalski \etal 1998, 2000, 2002, hereinafter referred to
as papers Ia, Ib, Ic) containing the mean values and errors of the
magnitude of all observed stars and their astrometry, we decided to release
the whole photometry data set, including all {\it I}-band epochs (HJD,
magnitude and its error) of over 40 million OGLE-II objects. To make the
data usable for the community, we created an additional database interface,
using the {\sc MySQL} database engine, allowing to make selections of
interesting objects based on any set of criteria on the object parameters
such as equatorial coordinates, mean and median magnitude, mean and median
error and the number of ``good'' points. The photometry is available in two
sets: the original Point Spread Function (PSF) photometry used during
OGLE-I and OGLE-II and the Difference Image Analysis (DIA) photometry which
is being used routinely in OGLE-III phase (Udalski 2003) and was used to
recalculate the photometry of the whole image set of OGLE-II. The database
system also includes the complete data set of the aforementioned {\it BVI}
maps, so the selection of the objects can also be based on the parameters
given in the maps.

The details of the structure of databases and the user interface to these
data is explained in the following sections of this paper. The project is
still under development. We plan to include the data obtained during OGLE-I
phase as well as data from the other, less frequently observed fields and
filters.
\vspace*{7pt}
\Section{Observations}
\vspace*{3pt}
The data being released come from the observations collected during the
second phase of the OGLE microlensing search conducted with the 1.3-m
Warsaw telescope at Las Campanas Observatory, Chile. The observatory is
operated by the Carnegie Institution of Washington. The telescope was
equipped with the camera built on a SITe ${2048\times2048}$ CCD
detector. The pixel size was 24~$\mu$m giving the 0.417~arcsec/pixel
scale. Observations were performed in the drift-scan mode, using the
``medium'' reading mode of the CCD detector (gain 7.1~e$^-$/ADU, readout
noise ${\approx6.3~{\rm e}^-}$) in the Galactic bulge fields and the
``slow'' mode (gain 3.8~e$^-$/ADU, readout noise ${\approx5.4~{\rm e}^-}$)
in the Magellanic Clouds. Details of the instrumentation setup can be found
in Udalski, Kubiak and Szyma{\'n}ski (1997).

The use of the drift-scan mode allowed to enlarge a single image to
${2048\times8192}$ pixels covering ${14.2\times57}$~arcmin on the sky.
Positions of some adjacent fields overlap by about one arcmin to allow 
the calibration tests. The effective exposure time in the {\it I}-band in
the Galactic bulge was 87~sec, increased to 99.3~sec after HJD\,=\,2451040
(Aug 14, 1998) and 125~sec in the Magellanic Clouds.
\vspace*{1pt}
\MakeTableee{|l|c|c|r|c|r|c|}{8cm}{OGLE-II fields and objects}{
\hline
\multicolumn{1}{|c|}{\uprule Target} &
\multicolumn{1}{|c|}{No. of} &
\multicolumn{1}{|c|}{Sky coverage} &
\multicolumn{1}{|c|}{No. of} &
\multicolumn{1}{|c|}{No. of} \\
                             &
\multicolumn{1}{|c|}{fields} &
\multicolumn{1}{|c|}{sq. deg.} &
\multicolumn{1}{|c|}{objects} &
\multicolumn{1}{|c|}{measurements\dorule} \\
\hline
\uprule
Gal. bulge& 49& 11 & 30.5$\times10^6$& 9.4$\times10^9$ \\
LMC       & 21& 4.5&  6.8$\times10^6$& 2.7$\times10^9$ \\
SMC       & 11& 2.4&  2.2$\times10^6$& 0.7$\times10^9$ \\
\hline}

Table~1 shows the number of fields, sky coverage, number of objects
detected and the total number of {\it I}-band 
measurements obtained by PSF photometry in
each of the main OGLE-II targets: Galactic bulge (GB) and the Magellanic
Clouds (LMC and SMC). The details regarding all these fields observed can
be found in Papers Ia,b,c.

\Section{Data Reduction and Calibration}
All the images taken from the telescope were reduced using the standard
OGLE data pipeline, described in detail in Paper~Ia. The PSF photometry of
the de-biased and flat-fielded frames was obtained using the modified {\sc
DoPhot} photometry program (Schechter, Saha and Mateo 1993) running in the
fixed position mode on sixty four ${512\times 512}$~pixel subframes. Each
frame was matched against the template image, obtained at very good seeing
conditions. Photometry of each subframe was tied to the photometry of the
template subframe by computing the mean shift derived from several hundreds
bright stars. Therefore, the photometry of the template image defines the
instrumental system for the PSF {\sc (DoPhot)} photometry. The
transformation of the instrumental photometry to the standard system was
calculated using several Landolt (1992) fields of standard stars observed
during about 250 photometric nights in the OGLE-II phase. For the technical
details and discussion of the transformation accuracy, refer to Paper~Ic.

The end of the OGLE-II phase coincided with the development of the
Difference Image Analysis (DIA) method of retrieving photometry, using an
image subtraction algorithm, especially useful in the dense stellar fields
(Alard and Lupton 1998 and Alard 2000). It was then implemented for the
OGLE-II Galactic bulge data by Wo\'zniak (2000) and the Magellanic Clouds
data (\.Zebru\'n, Soszy\'nski and Wo\'zniak 2001). The adopted DIA method
was used to create the catalog of OGLE-II microlensing events (Wo\'zniak
\etal 2001) and the general catalogs of candidate variable objects in the
Galactic bulge (Wo\'zniak \etal 2002) and the Magellanic Clouds (\.Zebru\'n
\etal 2001). The technical details of the DIA method and its OGLE
implementation are beyond the scope of this paper. The reader is referred
to the papers mentioned in this paragraph.

The successful introduction of the DIA method, yielding more accurate and
possibly deeper photometry of the OGLE-II data, convinced us to use this
method as the basic photometry tool in the third phase of the experiment,
OGLE-III. Further modifications to the DIA method allowed to measure not
only the variable objects but all the objects detected in the reference
image. This resulted in significant improving of the detection
effectiveness for faint objects (\eg RR~Lyr stars in SMC, Soszy\'nski \etal
2002) as well as low amplitude variables (\eg red giants in the Magellanic
Clouds, Soszy\'nski \etal 2004). This new approach, allowing to retrieve
the photometry of all objects, both variable and constant, prompted us to
apply the DIA method to recalculate the whole photometry of OGLE-II fields
using the standard OGLE-III pipeline (Udalski 2003). These data are now
being released to the public domain.

The DIA photometry provides only the difference of fluxes between the
current image and the reference image. To obtain absolute values, standard
PSF {\sc (DoPhot)} photometry of the DIA reference images was performed and
then tied to the well calibrated original OGLE-II photometry using mean
shifts computed on selected cross-identified objects.

Equatorial coordinates of all objects were calculated using third-order
transformation obtained by identification of several thousand bright stars
detected in OGLE fields in the Digitized Sky Survey images. The internal
accuracy of the determined equatorial coordinates, as measured in the
overlapping regions of neighboring fields is about
0\zdot\arcs15--0\zdot\arcs20. It is worth noting, however, that the
systematic error of the DSS coordinate system may reach 0\zdot\arcs7 and in
the extreme cases of objects close to the edges of the DSS images may even
exceed 1\arcs.

\Section{Photometry Database} 
This section briefly describes the core of the system -- the database
containing all photometric measurements of objects detected in the OGLE
fields, hereinafter referred to as {\sc PhotDB}. For each field, the {\sc
PhotDB} consists of four files:
\begin{enumerate}
\itemsep 0pt
\item frame index (DBI file), containing OGLE frame number, Heliocentric
Julian Date, exposure time, airmass, average FWHM in pixels and average
sky level, photometric grade of the frame, total of 40 bytes per frame.
\item object catalog (CAT), containing $X$, $Y$ coordinates on the template
frame, number of good measurements, as well as some internally used flags
and values, total of 36 bytes per object.
\item time index of frames (TI file), containing index allowing easy
chronological sorting of frames (which can be added to the database in
any order) without need of examining HJD of all frames, total of 2 bytes
per frame.
\item photometric data (previously DB or IDB, now SDB file, see below for
the explanation of formats), containing magnitude and its error value as
well as a packed flags byte, total of 5 (IDB, SDB) or 9 bytes (DB) per
measurement. The magnitudes values of all OGLE-II objects are corrected
for a small systematic error, caused by non-perfect flat-fielding at the
edges of the field (see Paper~Ic).
\end{enumerate}

The structure of {\sc PhotDB} is a modified version of the first photometry
database used in OGLE-I (Szyma\'nski and Udalski 1993). The main changes
include new formats of the biggest file of each field database, containing
the photometric data. In order to cope with the huge increase of the number
of measurements coming from the dedicated telescope on which the OGLE-II
phase was started (compared to OGLE-I) we had to make the databases smaller
by changing the format of the object magnitude and its error from 4-byte
float (DB files) to 2-byte short integer (IDB files) representing the
values in millimagnitudes. With the typical photometry error of 0.01~mag,
the retained accuracy was sufficient and we could compress the databases by
40\%. Even then, however, we could not keep all the sizes of the database
files below 2\,GB limit. Luckily, the new versions of operating systems
used ({\sc Solaris, Linux}) raised the maximum file size above this
threshold. Adopting the database software for the Large File Support (LFS)
was another important change.

The original formats of the photometry database files (both DB and IDB)
were designed to allow quick and easy adding new data coming from the
pipeline without complete rebuilding of the database. Thus, the new data
were simply appended to the database. As a result, the photometric
measurements of any given object were distributed sparsely all over the
database file. In order to make data retrieving more efficient the software
included an extensive system of memory buffers which significantly speeded
up the retrieval of the photometry of objects located close to each other
in the field catalog. However, for a single object or many objects selected
randomly, the access time was not negligible, especially as the databases
grew to the gigabyte sizes. For this reason another format was developed
for the databases which were already completed, or closed, after a season
of observations or the whole OGLE phase was ended. These databases contain
the photometry of all objects rewritten to form sequential, continuous data
chunks (SDB files). Now the retrieval of the photometric data of any given
object requires only one seek into the file and one read operation. The
time spent on these operations is insignificant compared to any reasonable
numerical analysis of the data.

\subsection{Good Photometric Points}
The idea of a ``good'' and ``bad'' photometric measurement has been evolving
since the beginning of the OGLE project, even though a {\sc bad\_phot} flag
was always present in the {\sc PhotDB} files. Unfortunately, the original
definition was too simple to be useful. Therefore in many OGLE publications
the authors introduced slightly different definitions. The new SDB
photometry files introduce a new, hopefully adequate method of estimating
the quality of a given measurement. The {\sc bad\_phot} flag is now set if
the individual magnitude error is bigger than 1.6 times ``typical'' error
for a given magnitude level (computed in 0.1~mag bins) for a given field
and filter. The ``absurd'' cases of negative or zero values of the
magnitude and/or error are also flagged.

\subsection{PSF \vs DIA Photometry Databases}
The DIA databases for the OGLE-II fields have a special status. They were
created well after the OGLE-II phase ended. The ``primary'' set of
photometric data for these fields has always been the PSF {\sc (DoPhot)}
photometry and the DIA photometry had to be cataloged consistently with the
existing PSF photometry databases. The most important result of this
approach is that although the DIA catalogs contain significantly more
objects than PSF catalogs, some of these additional objects may be spurious
detections or duplicate objects identified to more than one PSF
object. Some PSF objects have not been identified with any DIA object but
their (empty) entries remain in the catalog and the photometry files. All
these ``strange'' objects are properly flagged in the catalogs. The
recommended approach for a typical user of our databases is to make
selections of objects based on {\it BVI} maps or PSF photometry databases
and then to extract possibly more precise DIA photometry for these
objects. If, however, the user decides to query the DIA database directly,
we recommend to use ``no catalog flag'' option for the selection.

There is one particularly useful flag describing the individual
measurements in the DIA database, back-ported from the OGLE-III system. It
is set if the object is detected (independently of the information taken
from the reference image) in the subtracted frame. The presence of this
flag means that the object really did vary in that frame. The overall
number of measurements that are flagged this way is stored (as {\sf Ndetect})
in the catalog entry of every object. The selection criteria of database
objects may include a lower limit for the number of measurements flagged,
thus reducing significantly the possibility of selecting artificial
variables.

The low-level details of the structure of a DIA photometry catalog is
described in the Appendix.

\Section{Parameters Database}
To allow efficient retrieval of the photometric data we had to create an
interfa\-cing database containing basic parameters of all objects and a set
of tools to select objects fulfilling any set of criteria on these
parameters. We will hereinafter refer to this database as {\sc ParamDB}.
In order to create {\sc ParamDB} we have installed a {\sc MySQL} relational
database server which makes the tasks of creation, maintaining, updating
and searching the databases easy and efficient. There is a separate {\sc
ParamDB} for each main OGLE target (Galactic bulge, LMC, SMC), both for PSF
{\sc (DoPhot)} and DIA photometry. For each object in the database, the
following parameters are provided:
\begin{itemize}
\itemsep -2pt
\item {\hb Field} name (\eg BUL\_SC1)
\item {\hb StarID} -- object number in the {\sc PhotDB} of the Field
\item {\hb X,Y} -- coordinates of the Object on the template frame (in pixels)
\item {\hb StarCat} -- catalog designation: a string composed of {\it RA/Dec} in
the form of HHMMSS.SS$\pm$DDMMSS.S
\item {\hb RA, Decl} -- equatorial coordinates in hours, degrees
\item {\hb Ngood} -- number of good photometric points
\item {\hb Pgood} -- percentage of good photometric points
\item[--] All following parameters are calculated from the good points only.
\item {\hb Imean} -- mean {\it I} magnitude
\item {\hb Imed} -- median {\it I} magnitude
\item {\hb Isig} -- standard deviation of {\it I} magnitude
\item {\hb Imederr} -- median {\it I} magnitude error
\item {\hb Imin} -- minimum {\it I} magnitude
\item {\hb Imax} -- maximum {\it I} magnitude 
\item {\hb Ndetect} (DIA only) -- the number of independent detections in
the subtracted images (\cf Section~4.2).
\end{itemize}
Please note that the OGLE fields overlap slightly, so on the edges one can
find duplicate entries for the same objects.

The SQL database concept allows easy reconstruction of the databases. It is
quite possible that in the future we will add some new parameters to
facilitate selection of astronomically interesting objects. Feedback from
the database users will be appreciated.

\Section{{\bits BVI} Maps Parameters Database}
To facilitate selection of objects we have added to the {\sc MySQL} databases
the whole set of data released in previously published OGLE-II {\it BVI}
maps of the Magellanic Clouds and {\it VI} maps of the Galactic bulge.
The detailed description of the data set can be found in Papers Ia,b,c.
Here we only summarize the parameters included in this database,
hereinafter referred to as BVI-DB.
\begin{enumerate}
\itemsep -2pt
\item {\hb Field} name (\eg BUL\_SC1)
\item {\hb StarID} -- object number in the {\sc PhotDB} of the Field
\item {\hb RA, Decl} -- equatorial coordinates in hours, degrees
\item {\hb X,Y} -- coordinates of the Object on the template frame (in pixels)
\item {\hb V} -- mean {\it V} magnitude
\item {\hb V\,{\rm\bf --}\,I} -- mean $V-I$ color index 
\item {\hb I} -- mean {\it I} magnitude
\item {\hb Vgood} -- number of good photometric points in {\it V}
\item {\hb Vbad} -- number of bad photometric points in {\it V}
\item {\hb Vsig} -- standard deviation of {\it V} magnitude
\item {\hb Igood} -- number of good photometric points in {\it I}
\item {\hb Ibad} -- number of bad photometric points in {\it I}
\item {\hb Isig} -- standard deviation of {\it I} magnitude
\item[--] Following parameters are available only for the Magellanic Clouds
data
\item {\hb B\,{\rm\bf --}\,V} -- mean $B-V$ color index
\item {\hb B} -- mean {\it B} magnitude
\item {\hb Bgood} -- number of good photometric points in {\it B}
\item {\hb Bbad} -- number of bad photometric points in {\it B}
\item {\hb Bsig} -- standard deviation of {\it B} magnitude
\end{enumerate}

The BVI-DB database can be used not only to select objects to retrieve the
full photometry but also independently, as a useful tool to perform
statistical analysis of the OGLE-II {\it BVI} photometric maps.

\Section{User Interface}
The on-line access to the OGLE databases is provided through a simple WWW
interface available directly at {\it
http://ogledb.astrouw.edu.pl/\~{}ogle/photdb} or through a hyperlink from
the main OGLE WWW page {\it http://ogle.astrouw.edu.pl} or its US mirror
{\it http://bulge.princeton.edu/\~{}ogle}. The
underlying software consists of several PHP scripts using the low-level
{\sc PhotDB} utilities. The main, overview page contains the most important
information regarding the use of the interface. In the left panel, under
``Database Queries'', the user should choose which database parameter space
(``Photometric data'' or ``{\it BVI} maps'') to use for selections. The
appropriate Query Page is then loaded into the main panel.

\subsection{Making a Selection}
To make a selection of objects one should formulate a query using either
the Photometry ({\sc ParamDB}, \cf Section~5) or the {\it BVI} maps
parameters (BVI-DB, \cf Section~6) database. In both cases one has to
choose the target (Galactic bulge, LMC or SMC). The {\sc ParamDB} query
requires also selection of the photometry set (PSF or DIA).
\begin{figure}[htb]
\vglue-5pt
\centerline{\includegraphics[width=12.7cm, bb=5 160 610 780]{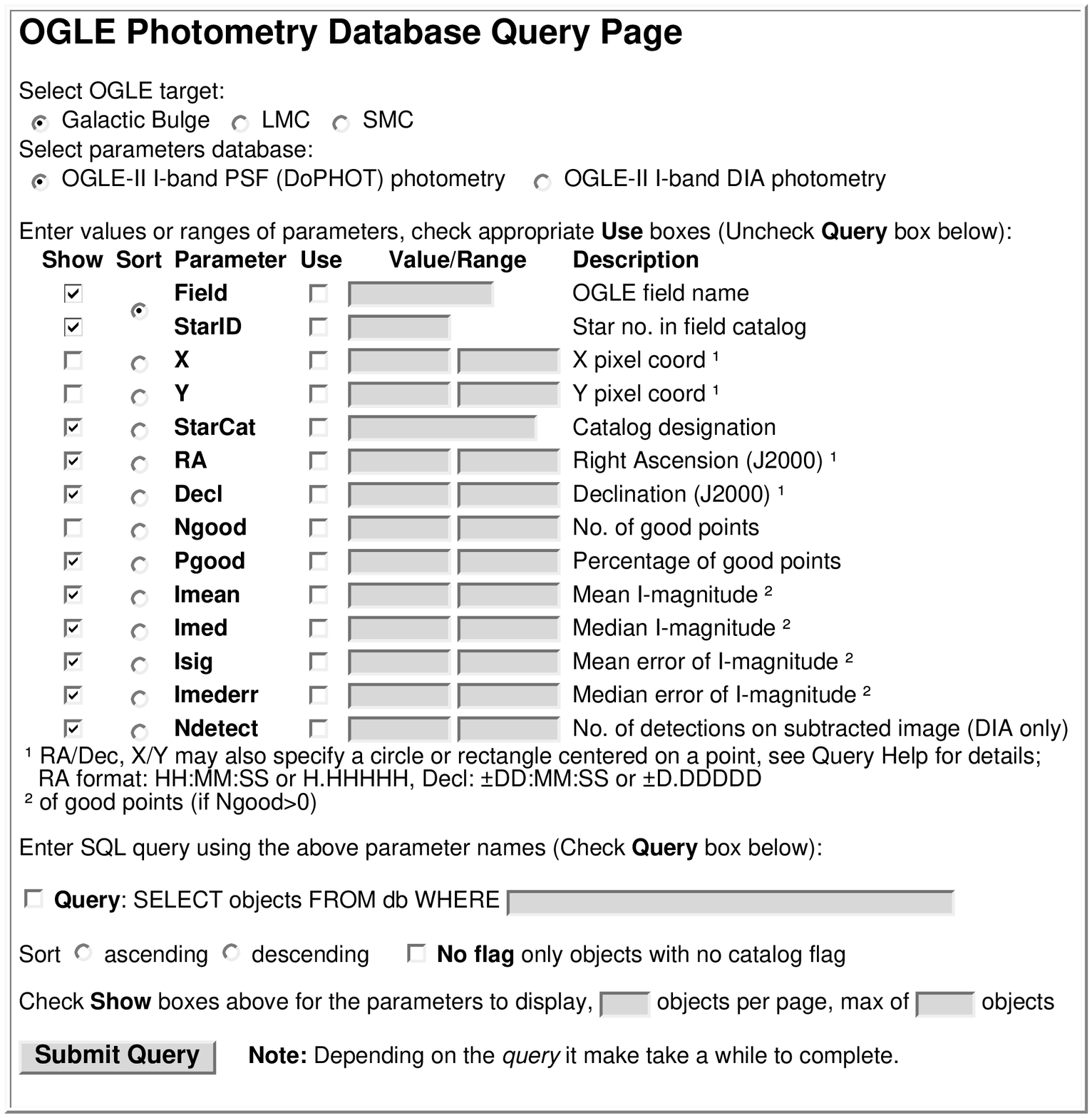}}
\vskip5pt
\FigCap{Photometry database query page.}
\end{figure}
\begin{figure}[htb]
\vglue-7pt
\centerline{\includegraphics[width=12.6cm, bb=5 95 610 780]{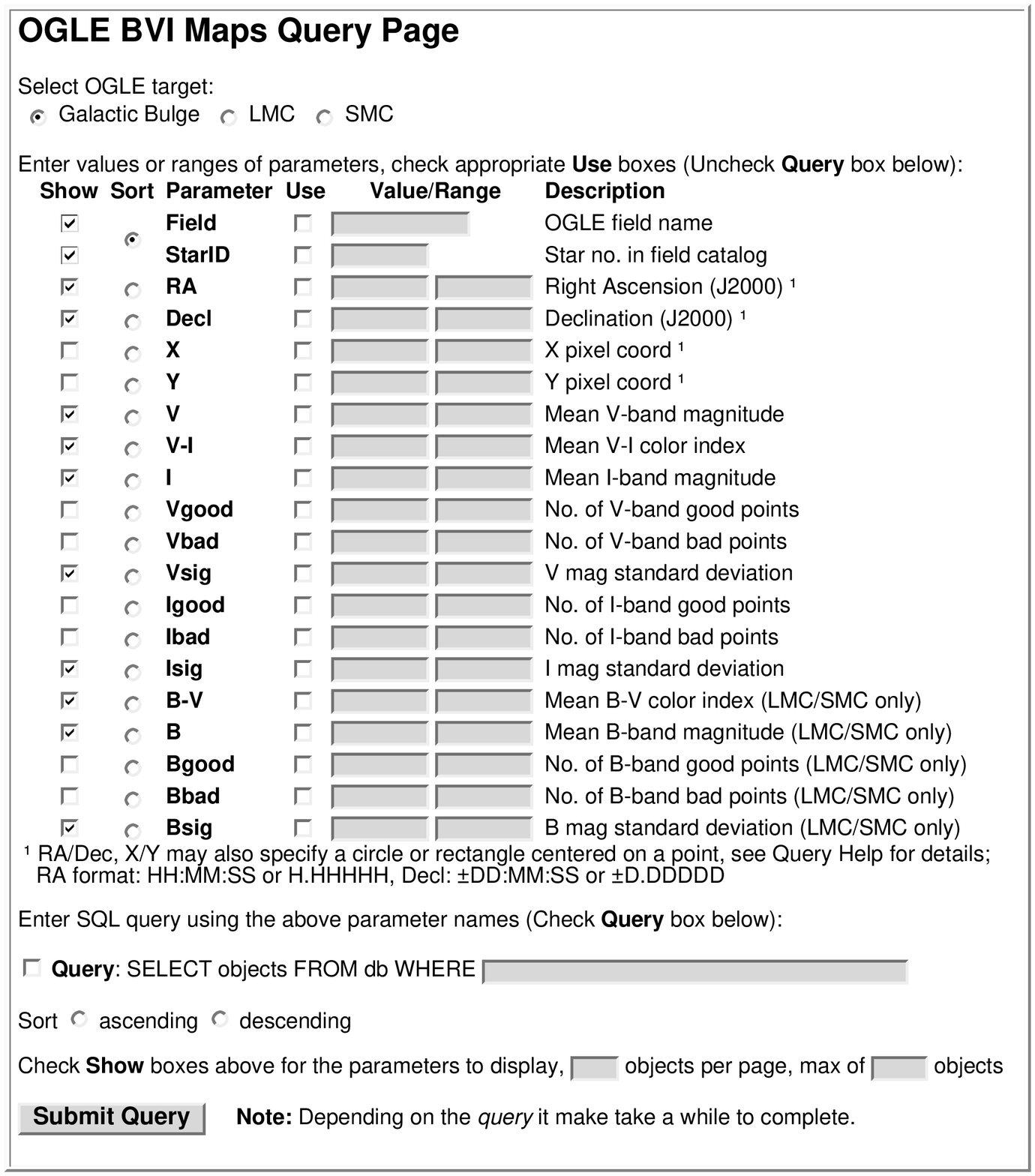}}
\vskip5pt
\FigCap{{\it BVI} maps database query page.}
\end{figure}
Each page ({\sc ParamDB}, BVI) is a {\sc html} form to be filled and
submitted to the server (Figs.~1 and~2). A selection of parameters can be
done in two possible ways:
\begin{enumerate}
\itemsep -1pt
\item By filling the input fields marked by relevant parameters names with
the minimum and maximum value of the parameter and checking the ``Use''
box. Leaving any of the limit values empty means no lower or upper limit
for the parameter value. All the used parameters limits are combined with
logical AND when formulating the query. The ``Query'' check box must be
unchecked.

Special rules work for coordinates ({\it RA/Dec}, {\it X/Y}): the range may be
specified in three ways: (a)~Normal min/max range values; (b)~A circle
around the point specified by ``left'' values. The circle radius (in arc
seconds for {\it RA/Dec}, in pixels for {\it X/Y}) should be given in one
(and only one) ``right'' value by prefixing it with letter 'r'; (c)~A
rectangle centered on the point specified by ``left'' values. Both
``right'' values should be given prefixed with 'r' with the meaning of the
rectangle half width/height.

\item If this simple logic is insufficient, one can manually enter the
WHERE part of SQL query to be submitted. An explanation of SQL syntax is
beyond the scope of this paper but it is an easy task. One can use the
parameter names, parentheses and relational operators (note for C
programmers: the equality operator is a single '='). It is also worth
noting that the parameter name for object declination is Decl, not
Dec (the latter form is a reserved SQL word). Short explanation
and a few examples are included on the Query Help page. The ``Query''
check box must be checked.
\end{enumerate}

Other input fields, boxes and buttons include:
\begin{itemize}
\itemsep -1pt
\item {\hb Show} check boxes located next to individual parameter names
control whether the values of these parameters are displayed in the result.
\item {\hb Sort} check boxes 
control which parameter will serve as the sorting key. The sort order can
be chosen by checking ``ascending'' or ``descending'' box at the bottom of
page. Note that the ``Field'' and ``StarID'' parameters constitute a single
sorting key.
\item {\hb No flag} check box: if set, only the objects having no catalog
flag set will be retrieved (excluded multiple identifications etc.,
see the Appendix)
\item {\hb Query} check box: if set, the SQL query manually inserted to the
adjacent input field will be used for the selection instead of parameter
value limits
\item {\hb Objects per page} input field sets the number of
objects to be displayed in one page of results.
\item {\hb Max objects} input field sets the maximum number of
objects to be selected.
\item {\hb Submit Query} button sends the query to the server for retrieval of
the list of objects fulfilling the given criteria.
\end{itemize}

\subsection{Reviewing the Results} 
If the query succeeds, the first page of results is displayed, containing
all the parameters chosen to be shown. Also, the total number of objects
found is shown and the hyperlinks to other pages if the result does not fit
into one page (see example in Fig.~3). Please note that the term ``page''
used here refers to the chunk of objects shown together and has nothing to
do with the actual browser window size.
\begin{figure}[htb]
\centerline{\includegraphics[width=13.3cm, bb=10 335 610 785]{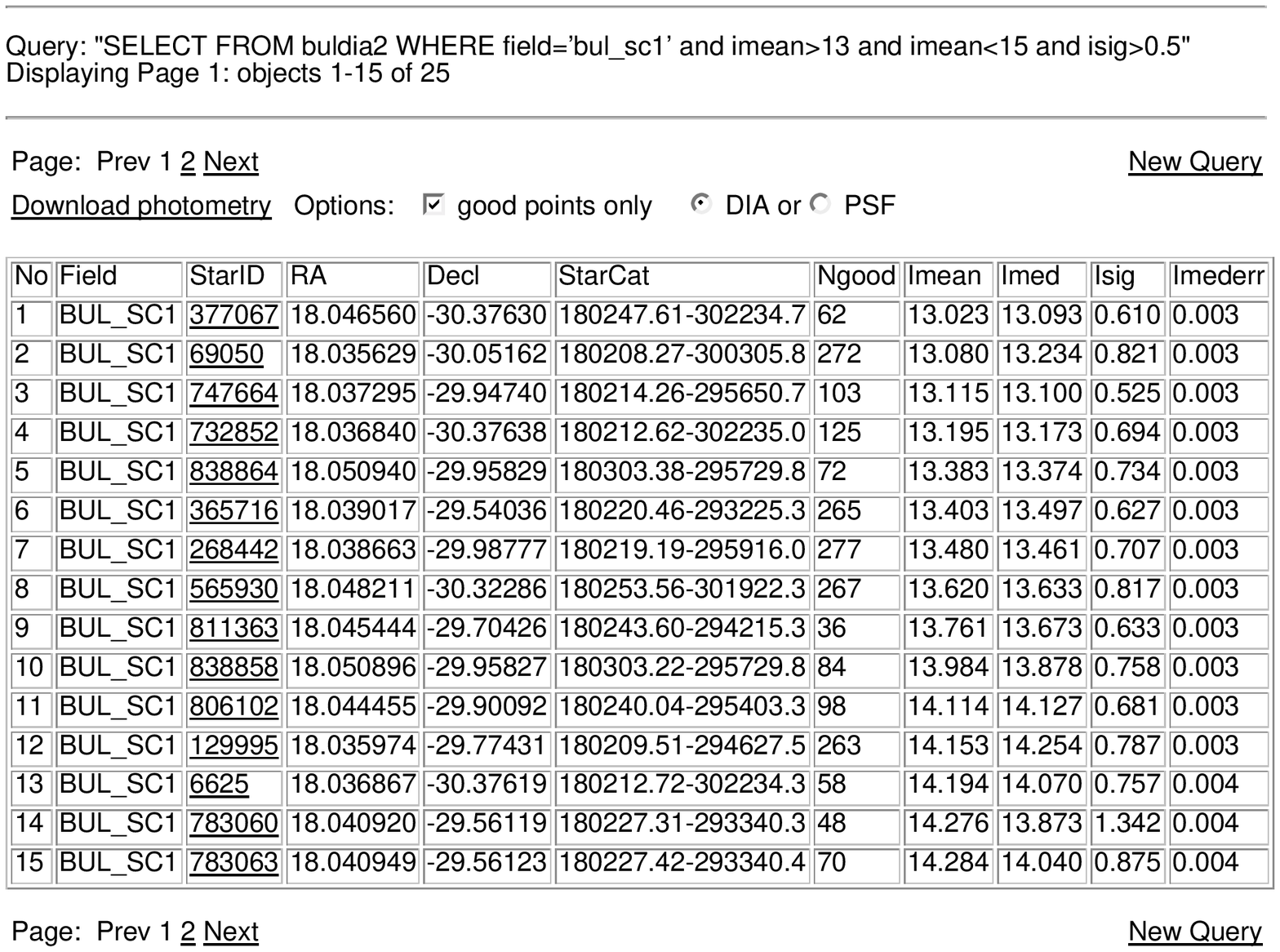}}
\vskip6pt
\FigCap{Query results page example.}
\end{figure}

Individual objects data can be displayed on a separate page by clicking on
the ``StarID'' number in the results table. The new page shows the
photometric parameters of the selected object, its light curve plot and the
photometric data (see example in Fig.~4). Both the light curve and the
photometry table can be downloaded by clicking appropriate hyperlink. The
next subsection describes the columns of the photometric data table.
\begin{figure}[htb]
\vglue-3mm
\includegraphics[width=13.5cm]{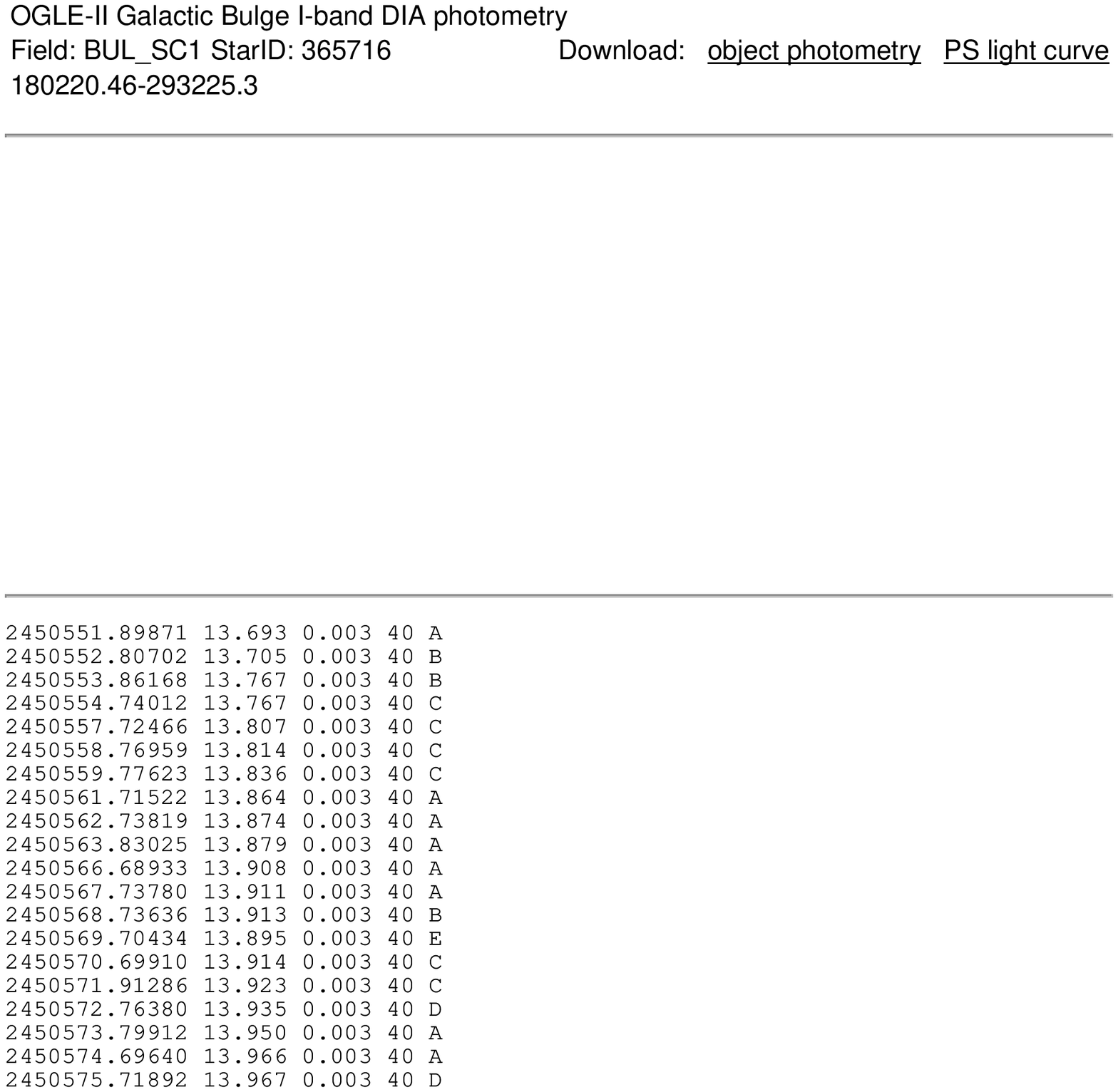}
\vglue -428pt
\hglue 25pt
\includegraphics[height=130pt, bb=70 70 460 390]{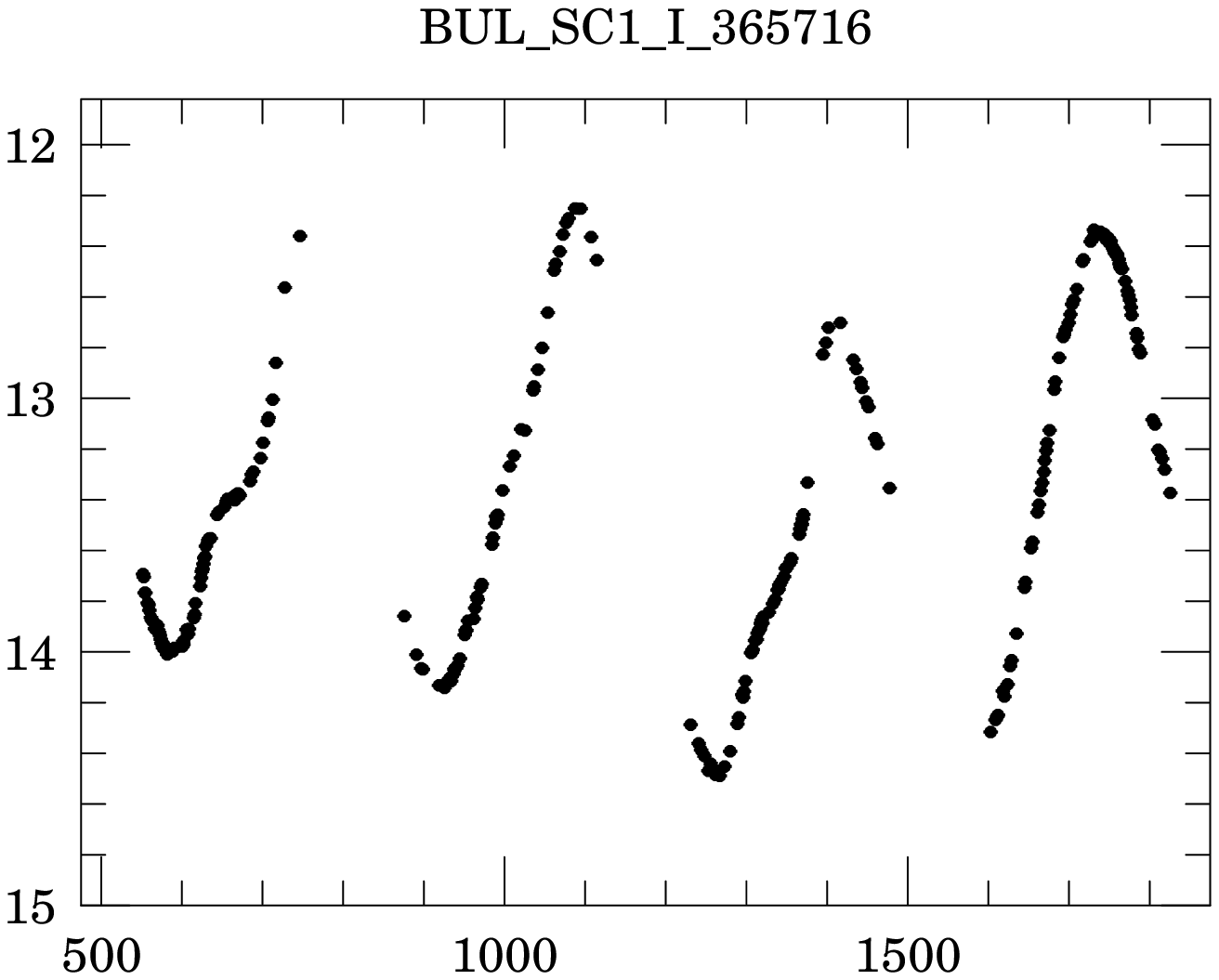}
\vglue 165pt
\FigCap{Individual object page example (only a sample of data points shown).}
\end{figure}

The photometry set used to retrieve the data (PSF or DIA) can be selected
by checking a box atop of any results table page. One can also choose to
include all points or the good points only.

\subsection{Downloading the Photometry}
The entire set of photometric data for all the objects selected by the query
can be downloaded by choosing the PSF {\sc (DoPhot)} or DIA data set and
clicking ``Download photometry'' link on the results page. The data are
delivered in the form of a gzipped tar file containing individual
photometric data files (named as {\sf Field\_StarID.dat}) containing 5
columns: HJD, magnitude, magnitude error, photometry flag (hexadecimal) and
the frame grade (A--F, best to worst). The flag contains combined catalog
flag for the object and the actual photometry flag of each measurement. For
the list of flags, their meaning and values, see the Appendix.

There is a limit of the number of objects for which the photometry can be
downloaded in a single query. It is currently set to 100\,000 but this
number may be changed in the future depending on the average data flow from
the server. Any query, however, will return the number of objects found.
The user can set a smaller limit for the query.

\Acknow{The author wishes to thank Drs.\ M.\ Kubiak and A.\ Udalski for
helpful discussions.
This paper was partly supported by the Polish KBN grant
2P03D02124. Partial support to the OGLE project was provided with the NSF
grant AST-0204908 and NASA grant NAG5-12212 to B.~Paczy\'nski. We
acknowledge usage of the Digitized Sky Survey which was produced at the
Space Telescope Science Institute based on photographic data obtained using
the UK Schmidt Telescope, operated by the Royal Observatory Edinburgh. We
also acknowledge usage of the {\sc MySQL} server kindly provided under the
free GPL license by {\sc MySQL} AB.}

\newpage
\begin{center}
{\bf Appendix: Technical Details of the DIA Photometry Catalogs}
\end{center}

\noindent Every DIA catalog (CAT) file has a following sequence of objects:

\begin{enumerate}
\itemsep -1pt
\item[(a)] objects cross-identified with {\it I}-band {\sc PSF (DoPhot)}
catalog
\item[(b)] all the remaining DIA objects (flagged in the catalog with
{\sc cat\_dia\_new} flag)
\end{enumerate}

\noindent The cross-identification here means ``the closest object
within 1.2 pixel (0\zdot\arcs5), if any''. This approach has the following
consequences:
\begin{enumerate}
\itemsep -1pt
\item The DIA catalog is always larger than the PSF {\sc (DoPhot)} catalog.
\item Some of the PSF objects (part~(a) of the DIA catalog) have no DIA 
object identified -- these objects are flagged in the catalog with {\sc
cat\_dia\_no\_phot} flag but both the catalog and (empty) photometry
entries are present to keep object numbers consistent.
\item Some DIA objects have been cross-identified with more than one PSF
DB objects. Entries for those objects are flagged with {\sc
cat\_dia\_mult\_doph} flag in the catalog and the corresponding photometry
entries are duplicated. The number of these ``multiplicities'' is placed in
{\sf nsame} field of catalog structure while {\sf same} field links all
such multiples (first $\rightarrow$ second $\rightarrow$ ... $\rightarrow$
last $\rightarrow$ first)
\item For some of PSF DB objects more than one DIA object has been
found within 1.2 pixel distance. These objects are flagged with {\sc
cat\_dia\_mult\_dia} in the catalog. If (and only if) they are not also in
category 3 above, {\sf nsame} field contains the number of DIA objects
within 1.2 pixel from a given PSF object and {\sf same} field contains the
number of the second closest object.
\end{enumerate}

\noindent The following additional notes further explain the details:
\begin{itemize}
\itemsep -1pt
\item all DIA catalog {\it X,Y} coordinates are taken from DIA photometry. The
only obvious exception are the "empty" entries (category 2 above) for which
the original PSF {\sc (DoPhot)} coordinates have been kept.
\item the above given categories 3 and 4 are NOT the same. In (3) the 1.2
pixel distance is counted from a PSF object position while in
(4) -- from a DIA object position. Even for cross-identified objects,
PSF and DIA {\it X,Y} coordinates may be slightly different.
\item all frames photometred by DIA have been included in the OGLE-II DIA
{\sc PhotDB}. Their grades (A--F, best to worst) are ``inherited'' from the
PSF {\sc (DoPhot)} {\sc PhotDB}. For those frames that were absent there,
grade 'F' is applied.
\item when the photometry is retrieved from the database for a given
object, the catalog flag for that object is combined with the flags of
each individual measurement. The following is a complete list of values
and meaning of the flags. The value listed in the downloaded photometry
files is a bitwise or-ed sum of the individual flags. Please note that
some of the flags were used in OGLE-I phase only (for which the
photometry is not yet available on-line) but they are given here for
completeness.
\end{itemize}
\begin{tabular}{lcp{215pt}}
{\sc cat\_subf\_mult} & 0x01 & multiple detection on subframe (OGLE-I
only) \\
{\sc cat\_edge\_obj} & 0x02 & object close to the frame edge (OGLE-I
only) \\
{\sc dia\_mult} & 0x10 & DIA database object has any of the above
\newline described multiplicity flag set \\
{\sc bad\_phot} & 0x20 & the measurement was classified as bad \newline 
(\cf Section 4.1) \\
{\sc dia\_detected} & 0x40 & DIA database object was independently
measured on the subtracted image (\cf Section 4.2)
\end{tabular}
\end{document}